# X-ray Raman scattering with Bragg diffraction in a La-based superlattice

J.-M. André*, P. Jonnard, C. Bonnelle

*Laboratoire de Chimie Physique-Matière et Rayonnement, Université Pierre et Marie Curie LCP-MR , UMR-CNRS 7614, 11 rue Pierre et Marie Curie, 75231 Paris Cedex 05, France*

E. O. Filatova

*Institute of Physics, St Petersburg University, Ulianovskaya 1, St Petersburg, 198904, Russia*

C. Michaelsen, J. Wiesmann

*Incoatec GmbH, Max-Planck-Strasse 2, 21502 Geestacht, Germany*

**Abstract**

The non-dispersed soft x-ray emission from a La/$B_4C$ periodic multilayer irradiated by monochromatic x-rays has been measured as a function of the incident photon energy in the 125-200 eV range for different scattering angles. We have observed a scattered intensity peak at incident energies which shift towards the low-energy side as the value of the scattering angle increases. These observations are interpreted as Raman scattering by the 5p level of lanthanum assisted by Bragg diffraction from the lattice of the periodic multilayer. A simple model based upon the laws of energy and momentum conservation enables us to account for the variation of the incident energy associated with the Bragg-Raman peak.

## 1. Introduction

Periodic multilayer structures with layer thicknesses in the nanometer range, labelled hereafter superlattices, can be regarded as crystalline systems of adjustable interplanar distance. These structures are now widely used for optical and spectroscopic purposes in the soft-x-ray domain [1,2]. Due to both atomic character of their constituents and their periodic arrangement, the superlattices offer the possibility to observe atomic processes such as fluorescence, Rayleigh or Raman scattering, in combination with optical phenomena such as specular or Bragg reflection.

In this paper, we report on the interaction of a La-based superlattice with a monochromatic radiation in the La 4d threshold region. The relative intensity of non-dispersed scattered radiation is measured as a function of the *incident* photon energy and of the scattering angle. We propose to interpret the measurements by a combination of an atomic process with the Bragg reflection by the planes of the periodic structure.

During the eighteens, Raman scattering and fluorescence in stratified media were extensively studied both experimentally and theoretically in the optical domain [3-5]. Since X-ray Raman scattering has been evidenced [6-8], resonant inelastic x-ray scattering (RIXS) has received a lot of attention, in connection with the development of third generation synchrotrons [9-11]. Inelastic scattering requires a treatment up to the second order in perturbation theory leading to the so-called Kramers-Kronig formula with an explicit expression of the momentum conservation through the Kronecker delta [12]. This law of conservation has been used to investigate the band structure of crystals by x-ray Bloch **k**-selective resonant inelastic scattering [13,14]. In this case, the momentum transfer takes place to one electron-hole pair present at the final state in the valence and conduction states of the solid *modulo* any reciprocal-lattice vector **G** but no Bragg diffraction of the scattered photon is achieved. The phenomenon reported in the present work is attributed to a Raman scattering process assisted by a full momentum transfer of a reciprocal-lattice vector G.

## 2. Experiment and results

A La/$B_4$C superlattice with 20 bilayers was fabricated using a diode sputtering technique [15]. The thickness of the La and $B_4$C layers measured by grazing incidence x-ray reflectometry are respectively 4.69 nm and 4.87 nm, so that the superlattice period is 9.56 nm.

The superlattice was irradiated by the monochromatic radiation supplied by the 1200 lines/mm plane grating monochromator associated with the undulator of the UE56/1-PGM-b beamline at the BESSY synchrotron facility. The spectral bandwidth of the quasi-monochromatic radiation delivered by the monochromatic is close to 100meV and the radiation is circularly polarized. The geometry of the experiment is given in Figure 1. To carry out our measurements, we used the polarimeter-goniometer developed by Schäfers et al. [16]. Only the goniometer of the apparatus was used. The value of the glancing angle $q_{in}$ is equal to 31°. Let us note that in this condition, the energy diffracted according to the standard Bragg reflection, that is in the specular condition, would be close to 126 eV when the refractive index correction is not taken into account. The radiation is detected by means of a GaAs diode connected to a Keithley picoammeter. The 100 mm entrance slit located in front of the diode at 100 mm from the superlattice gives a 1 mrad divergence.

The scattering angle $q_{out}$ is defined as the angle between the direction of the incident beam and the direction of observation. For different values of $q_{out}$, the emitted intensity was measured versus the incident energy $E_{in}$. The results of these measurements are called hereafter « excitation spectra ». Let us emphasize that the emitted radiation is not dispersed. The incident energy $E_{in}$ has been varied in a spectral domain going over the La 4d excitation transitions and the La 4d ionization threshold. Experiments were also performed for $LaF_3$ and $La_2O_3$ evaporated films for comparison with the superlattice. The films were prepared by thermal evaporation of high purity powder ; $LaF_3$ was deposited on a silicon substrate and $La_2O_3$ on a silica glass substrate.

For the sake of clarity, we restrict this communication to the spectral domain (125-200 eV), above the La 4d ionization threshold, that is the range where a process characteristic of the superlattice has been observed. The whole spectra covering the La 4d ionization will be discussed in detail in a forthcoming paper.

The curves giving the emitted intensity from the superlattice versus the incident energy, that is the excitation spectra, are shown in Figure 2 for different scattering angles $q_{out}$ ranging from 43 to 67°. The relative intensity variation between each curve is not precisely known, then the spectra are normalized with respect to their maximum. One observes a peak which shifts towards the low-energy side as the value of the scattering angle $q_{out}$ increases and then which remains around 120 eV for $q_{out}$ greater than 85°. The amplitude of this peak increases as the scattering angle $q_{out}$ becomes close to the specular angle, that is $q_{out}= 62°$. The energy position of the maximum of the peak $E_{max}$ is plotted versus $q_{out}$ in Figure 3. We do not observe such a moving peak with the two lanthanum compounds.

**3. Discussion**

The lanthanum 4d ionization energy in the superlattice is in the 106-109 eV range. The most probable radiative transition from the La $4d^9$ ion takes place with emission of the $4d^9$-$5p^5$ electric dipole line (cf. Figure 4a). The La 4d spectrum presents a peculiarity : one of the 4d-4f excitation lines is located above the ionization threshold, at about 115-118 eV, and it is the most intense transition in this energy range [17,18]. The corresponding excited state, $4d^9 4f^1$ $^1P$, around 3 eV wide, has a large probability to decay by autoionisation to the $4d^9$ state (cf. Figure 4b). Then the radiative decay from this state takes place mainly with emission of the $4d^9$-$5p^5$ line. Other radiative decays occur but they can be neglected here.

In a RIXS process, the difference between the incident and scattered photon energies is equal to the difference between the initial and final states of the atoms in the considered process. For RIXS taking place in the energy range from 106-109 eV, one expects the energy difference to be equal to the energy labelled $E_x$, of the La $5p^5$ configuration. The value of $E_x$ is estimated to be about 20 eV in the superlattice [18,19].

To describe from a kinematical point of view the Raman process associated with Bragg diffraction, we write the laws of energy and momentum conservation for the whole system : photons, atoms and multilayer structure. It gives:

$$E_{in} = E_{out} + E_X \qquad (1)$$

and

$$\hbar \mathbf{k}_{in} = \hbar \mathbf{k}_{out} + \mathbf{q}_X + \hbar \mathbf{G}, \qquad (2)$$

where $E_{in}$ and $E_{out}$ are the energies of the incident and scattered photons respectively, $\hbar \mathbf{k}_{in}$ and $\hbar \mathbf{k}_{out}$ the momenta associated with the incident and scattered photons, and $\mathbf{q}_x$ the momentum associated to the atomic state X of energy $E_x$. Combining these two equations together with the dispersion equations which relate the energy to the momentum, leads to the following relationship which links the value of the resonant excitation energy $E_{in,max}$ to the scattering angle $q_{out}$:

$$E_{in,max} = \frac{E_X}{2} + \frac{\sqrt{2c^2 q^2 (1-\cos q_{out}) - E_X^2 \sin^2 q_{out}}}{2(1-\cos q_{out})} \tag{3}$$

with

$$q^2 = (\mathbf{q}_x + \hbar \mathbf{G})^2 = q_x^2 + (\hbar G)^2 + 2 q_x \hbar G \cos(q_X), \tag{4}$$

where $q_x$ is the angle between the direction of $\mathbf{G}$ and $\mathbf{q}_X$.

This equation can be regarded a Bragg law generalized to the case of inelastic scattering. Indeed, simple calculations allow one to check that the standard Bragg law which is valid for elastic scattering ($E_{in} = E_{out}$) in specular condition ($q_{out} = 2q_{in}$) is recovered for the case $E_x = 0$.

Our experimental results can be directly compared to the values given by Eqs. (3) and (4) with $E_x = 19.5$ eV, $G = G_1$ and $q_x$ being calculated by assuming that the direction of $\mathbf{q}_x$ is the direction of observation, i.e. $q_x = \frac{p}{2} + q_{out} - q_{in}$. The value of $G_1$ has been calculated by taking into account the refraction correction to the Bragg diffraction at the first order in the expansion in $d$ [1,2]. This value is given by::

$$G_1 = 2p \left[ d \left( 1 - \frac{d}{\sin^2 q_{out}} \right)^{-1} \right], \tag{5}$$

where d is the geometric thickness of the bilayer (9.56 nm) and $d$ is the unit decrement of the real part of the refractive index, which depends on the photon energy of the outgoing photon. The values of $d$ can be found in [20,21]. The index correction changes the value of Bragg energy from 126 to 142 eV for an incidence angle of 31°. The present correction is large because the involved energy is in the La 4d anomalous scattering region.

Our assumption on $q_x$ allows us to find the best fit between theory and experiment. Let us emphasize that our simple model does not allow to justify this assumption. It is interesting to compare the position of the "resonant" excitation energy $E_{in,max}$ with the Bragg energy calculated with a standard code [22] at the angle $q_{out}$ in the specular condition. This is done in Figure 3. It appears that neither the data of the above model neither the experimental results agree with the Bragg energy calculated in the framework of the standard Bragg diffraction. Indeed the relevant Bragg law corresponding to the phenomenon that we observe is given by a Bragg formula generalized for inelastic scattering as the one given by Eqs. (3) and (4). A dynamical model using a theoretical approach similar to the one given in [23] but extended to the inelastic scattering should be implemented to account more accurately for our experimental results. This work is in progress.

## 4. Conclusion and perspectives

X-ray Raman scattering assisted by Bragg scattering has been observed in a La-based superlattice. A simple model based upon the laws of energy and momentum conservation enables to account for the variation of the "resonant" excitation energy as a function of the experimental geometry. One expected that Bragg diffraction gives rise to an enhancement of the scattering process but further experiments are needed to check this point. Measurements of the spectrum of the emitted radiation are planed in a low-resolution mode using a multilayer mirror as analyser, especially to check this enhancement. The determination of the polarization of the scattered radiation would be also relevant. Unfortunately, this task will be difficult to perform because of the low scattered intensity and the poor efficiency of polarimeters in the soft x-ray range.

It appears from this work that the scattering cross-section is sensitive to non-local properties of the photon field ; far from being only determined by the properties of the interacting atoms, the scattering process may be strongly modified by the structure of the medium which controls the boundary conditions for the scattered photon field.

This work suggests that it should be possible to control the intensity of Raman scattering in the soft-x-ray domain, as it was done in the optical domain [24].

**Acknowledgments :** This work was partly carried out at BESSY in the framework of the project Number BI3A-21/300804 and supported through the European Commission under I 3 Contract 3-CT-2004-506008. The authors thank Drs. A. Gaupp and F. Schäfers from BESSY for their help during the measurements.

**Figure captions :**

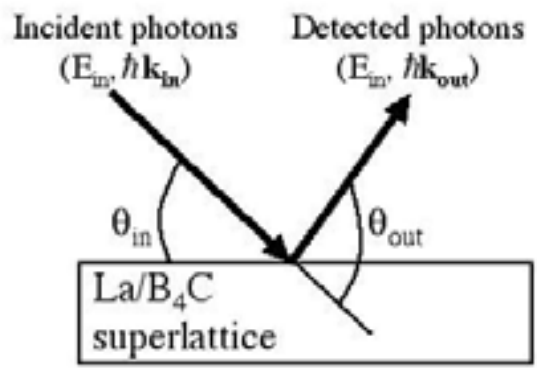

Figure 1 : Geometry of the experiment.

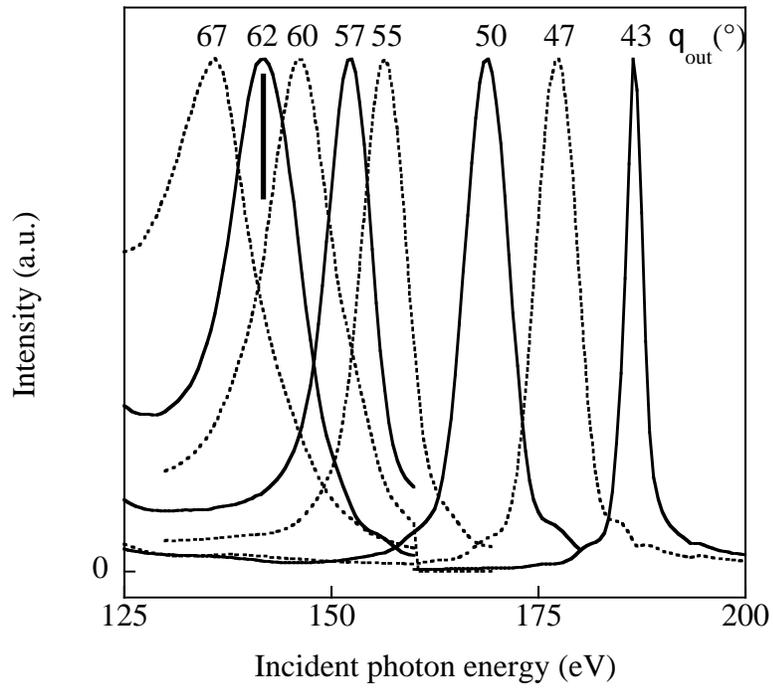

Figure 2 : Excitation spectra of the La/B$_4$C superlattice for different scattering angles ranging from 43 to 67 degrees. The incidence angle is 31°. The spectra are normalized with respect to their maximum. The vertical bar marks the specular reflected beam

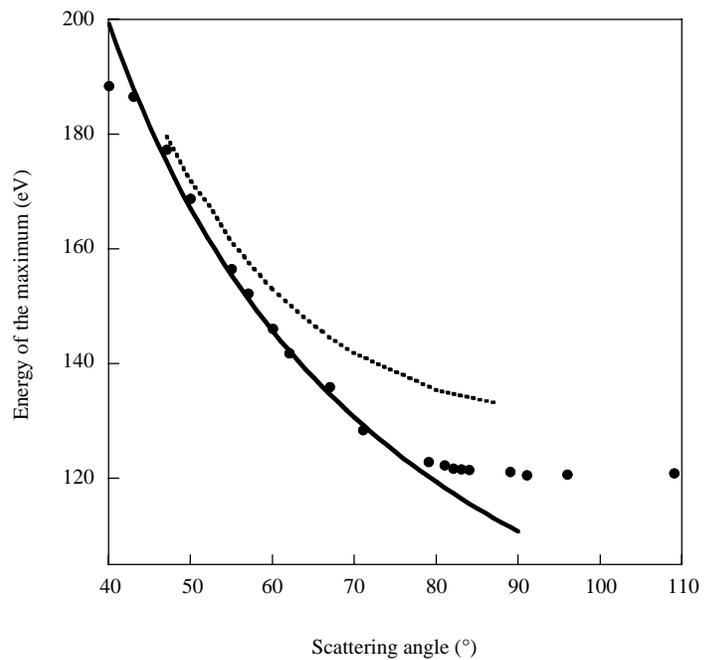

Figure 3: Peak position of the "resonant" excitation energy versus scattering angle : experiment (dots), our model (solid line), Bragg energy (dotted line).

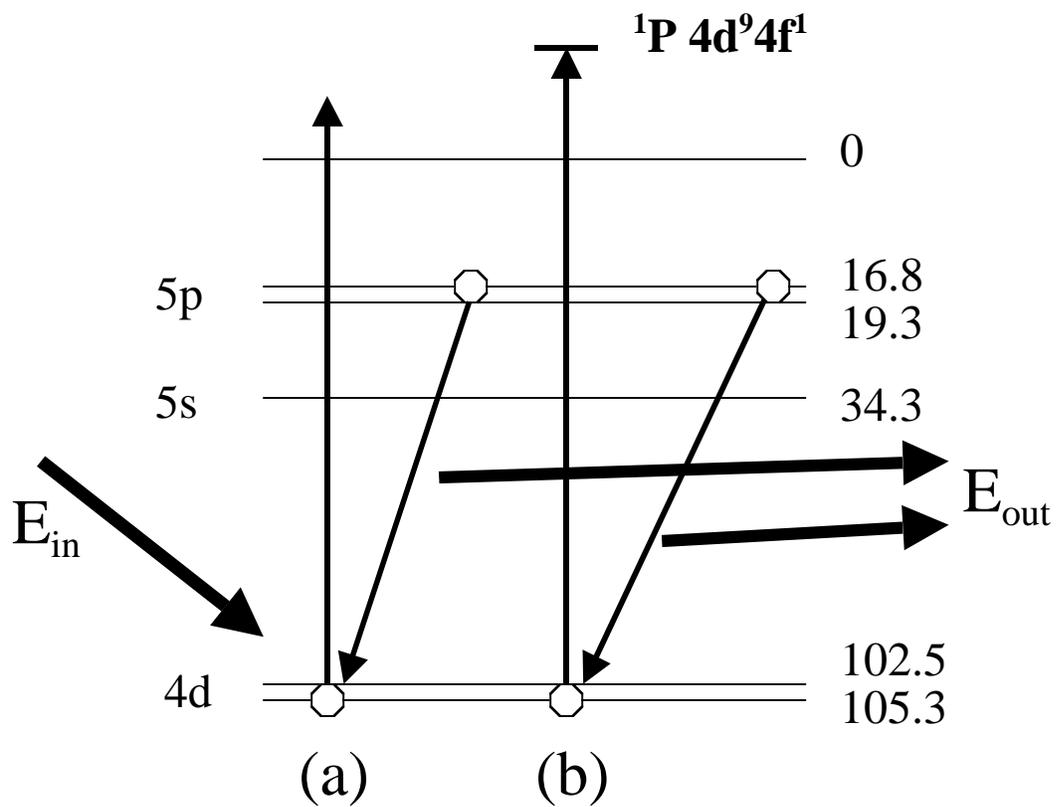

Figure 4 : Lanthanum energy levels (in eV) and scattering processes involving (a) the continuum and (b) the autoionising ¹P state.